\documentclass[aip]{revtex4-1}
\usepackage{graphicx}
\usepackage{bm}
\usepackage{fancyhdr}
\usepackage{amsmath}
\usepackage{amsfonts}
\usepackage{amsbsy}
\usepackage{amssymb}
\usepackage[mathscr]{eucal}
\usepackage{color}
\usepackage{soul}
\usepackage{mathtools}
\begin{document}
\title{Static vs dynamic rough energy landscapes: Where is diffusion faster?
}
\author{Dmitrii E. Makarov}
\email{makarov@cm.utexas.edu}
\affiliation{Department of Chemistry and Oden Institute for Computational Engineering and Sciences, University of Texas at Austin, Austin, TX, 78712 }
\author{Peter Sollich}
\email{peter.sollich@uni-goettingen.de}
\affiliation{Institute for Theoretical Physics, Georg-August-Universität Göttingen, 37077, Göttingen, Germany}

\begin{abstract}
Molecules in dense environments, such as biological cells,  are subjected to forces that fluctuate both in time and in space.  While spatial fluctuations are captured by Lifson-Jackson-Zwanzig's model of ``diffusion in a rough potential'', and temporal fluctuations are often viewed as leading to additional friction effects, a unified view where the environment fluctuates both in time and in space is currently lacking. Here we introduce a discrete-state model of a landscape fluctuating both in time and in space. Importantly, the model accounts for the back-reaction of the diffusing particle on the landscape. As a result we find, surprisingly, that many features of the observable dynamics do not depend on the temporal fluctuation timescales and are already captured by the model of diffusion in a rough potential, even though this assumes a static energy landscape.   
\end{abstract}

\maketitle

\section{Introduction}
In many problems concerned with condensed-phase systems, one is interested in the dynamics of a selected degree of freedom $x$ coupled to other fluctuating degrees of freedom.  For example, single-molecule measurements \cite{makarov2015single} and single-particle tracking experiments\cite{single_tracking, Fazel_Grussmayer_Ferdman_Radenovic_Shechtman_Enderlein_Pressé_2024}  inherently probe low-dimensional observables.  While a formally exact theory exists allowing one to project out unobserved degrees of freedom if the underlying microscopic dynamics is known\cite{Zwanzig,mori1965transport,grabert2006projection},  it often leads to equations of motion that are intractable and/or difficult to relate to experimental observables \cite{Glatzel_Schilling_2021,hijon2010mori,Ayaz2022},  and a key question is how to model the effect of the  fluctuating environment in a feasible way. Many practical approaches to this problem are based on the view that the environment alters (usually reduces) the effective diffusivity of the observed degree of freedom. Two models are often invoked to quantify this effect. The first, originally studied by Lifson and Jackson \cite{Lifson_Jackson_1962} and later generalized by Zwanzig \cite{Zwanzig_1988} and  by  others \cite{Lifson_Jackson_1962,De_Gennes_1975,Zwanzig_1988,Bryngelson_Wolynes_1989,Gray_2020} recognizes that other degrees of freedom create a rough, {\it spatially} fluctuating energy landscape. As the system diffusing along $x$ repeatedly gets trapped in local energy minima, the effective diffusivity observed over lengthscales larger than that of the lengthscale of the landscape fluctuations becomes renormalized with a factor that depends on the statistical properties of the fluctuating energy landscape. 

The second model recognizes that if $x$ is subjected to a force that fluctuates {\it in time}, then the fluctuation-dissipation theorem leads to an associated friction force. Dielectric friction models\cite{Hubbard_Onsager_1977, Nee_Zwanzig_1970,Samanta_Matyushov_2021,Hagen_electric_friction} and certain simulation-derived Langevin equation models\cite{Straub1986,Straub_1987, Netz2025}  are of this type. 
The dielectric friction models, in particular, account for the fact that the coupling of a charged object (e.g., a molecule) to the electric field induced by the environment leads to an additional energy dissipation mechanism. We note that this type of models can be only derived from first principles in special cases\cite{Zwanzig,MakarovLangevin2013,elber2020molecular}.  

The predictions made by the two types of models differ qualitatively: for example,  landscape roughness has a multiplicative effect on the apparent diffusivity and thus on the apparent friction coefficient\cite{Zwanzig_1988}  (see below), while dielectric friction is an additive term  in the total friction coefficient, at least assuming that solvent and dielectric forces are decoupled\cite{Matyushov_2024}.  Such differences are related to the underlying  physical assumptions:  in the first model, the energy landscape is frozen, while in the second it fluctuates in time. We expect that both models should arise as limiting cases of a more general model with energy landscape fluctuating {\it both in space and in time}. Indeed, the results of Lifson--Jackson,  and Zwanzig's theories  can be recovered as a quenched disorder limit within a model of a tracer particle coupled to a random field\cite{Dean2007-bq,Dean2011-up}.  Estimating an effective diffusivity in a general case for such a model is challenging, however, and  studies have so far been limited to computer simulations and to perturbative treatments\cite{Dean2007-bq, Dean2011-up,Demery2011-ak}. As a  step towards a more general theory, here we study a  discrete model of a random walker interacting with a dynamically fluctuating lattice landscape and show that certain properties of the walk only depend on the strength of the interaction between the walker and  the landscape, but not on the timescale of landscape fluctuations. 

\begin{figure}
    \centering
    \includegraphics[width=0.75\linewidth]{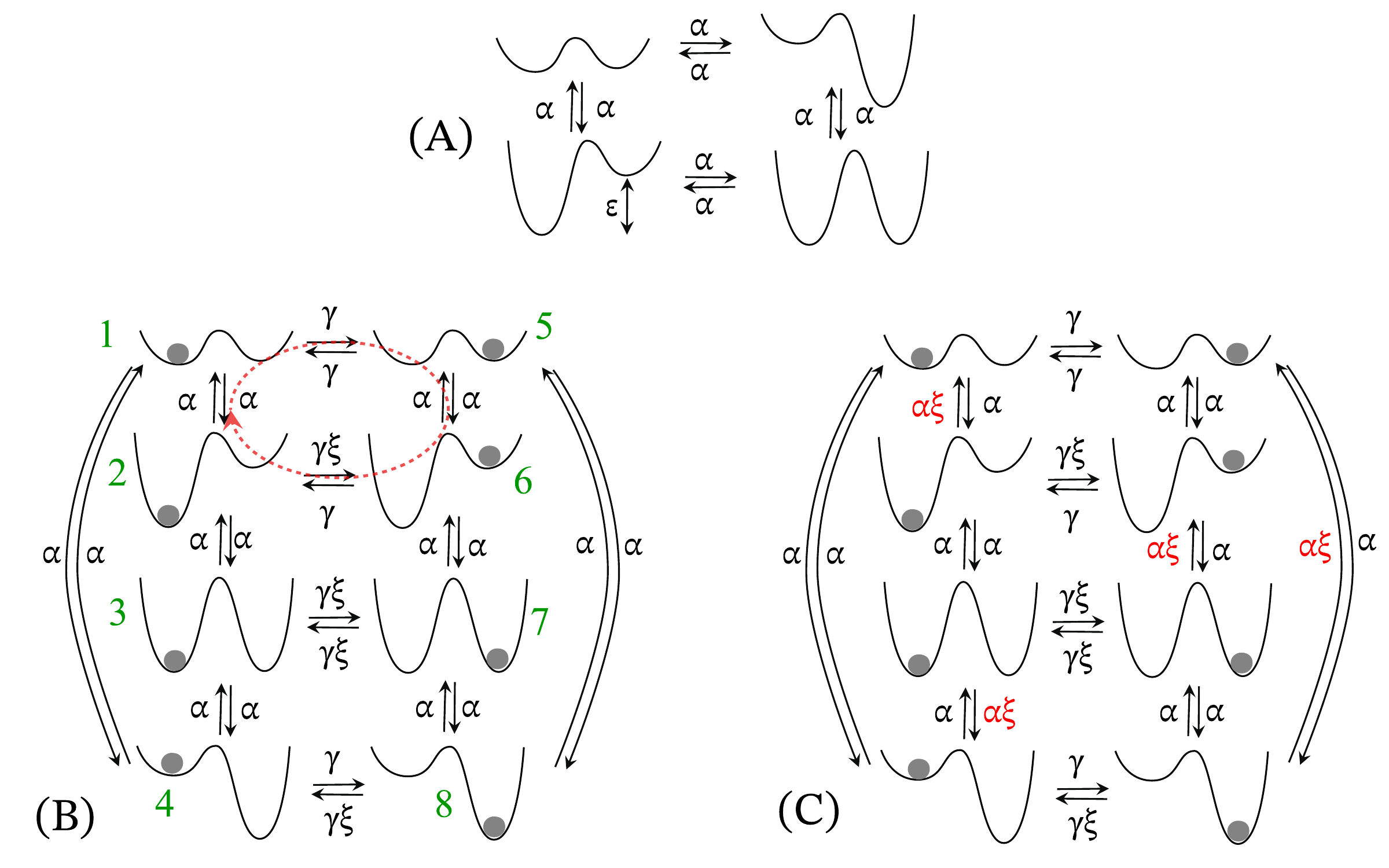}
    \caption{Toy model of a random walker in a fluctuating two-well landscape. (A): In the absence of the walker each potential well switches between ``deep'' and ``shallow'' states with a rate $\alpha$. (B): When the walker escapes the shallow well with a rate $\gamma$ and the deep well with a rate $\gamma \xi$, the kinetics of the system can be described by a kinetic network with eight states, which are shown and enumerated here. Transition rates are also shown, e.g., $k_{1 \to 5}=\gamma$, $k_{1 \to 2}=\alpha$. This network violates detailed balance, with an example of a 4-state dissipative cycle shown as a dashed oval. (C):  The transition rates of the network shown in (B) are modified such that the network satisfies detailed balance. The modified rates are highlighted in red.}
    \label{fig:two-well}
\end{figure}

This finding is somewhat surprising: starting with a frozen landscape limit,  it seems that the dynamics along $x$ will be always accelerated when the landscape is allowed to fluctuate in time, as the barriers trapping the walker will be lowered from time to time, allowing it to escape the traps more frequently (Fig.~\ref{fig:two-well}). The central finding of this paper is that this conclusion is not necessarily correct because detailed balance necessitates that the walker alters the dynamics of the landscape itself. To illustrate this we start with a simple toy model, in which a random walker jumps between two potential wells whose depths can fluctuate, independently, between ``deep'' and ``shallow'' states (Fig.~\ref{fig:two-well}). In the absence of the walker, each well switches between the two states with a rate $\alpha$, as shown in Fig.~\ref{fig:two-well}A. Suppose the rates of the walker's escape from the shallow and the deep wells are $\gamma$ and $\gamma \xi$, where $\xi=e^{-\beta\epsilon}<1$ ($\beta$ being the inverse temperature) is determined by the depth $\epsilon$ of the deeper well relative to that of the shallow one (Fig.~\ref{fig:two-well}A). If the rates of transitions between the deep and shallow states are unaffected by the walker, as shown in Fig.~\ref{fig:two-well}B, then the resulting system violates detailed balance: indeed, the full kinetic network describing all possible transitions between various states of the system contains detailed-balance-violating dissipative cycles (i.e.\ cycles in which the products of the forward rate coefficients are different from those of the backward rate coefficients, resulting in overall circular motion and thus in violation of time-reversal symmetry), one of which is shown in Fig.~\ref{fig:two-well}B. In order to satisfy detailed balance, the rates of transitions between certain well states must be modified, as illustrated in Fig.~\ref{fig:two-well}C. Physically, if the walker is ``stabilized'' by the interaction with its environment resulting in a longer lifetime in a deeper potential well, the ``deep well'' state of the environment is itself stabilized by the walker. 

We now wish to calculate the mean time $\tau$  that the system dwells on the left (identical to the mean dwell time on the right in view of the symmetry), and, in particular, its dependence on the parameter $\alpha$ characterizing the timescale of the landscape fluctuations. The inverse of this time is the average frequency of switching from left to right, $\tau^{-1}=k_{L\to R} =k_{R\to L} $, which can be calculated as the equilibrium unidirectional flux from left to right divided by the probability to be on the left,
\begin{equation} \label{flux-over-pop}
    k_{L\to R}=\frac{p_1 k_{1\to 5} +p_2 k_{2\to 6} +p_3 k_{3\to 7}+p_4 k_{4\to 6}}{p_1+p_2+p_3+p_4}. 
\end{equation}
Here $p_i$  is the steady-state probability to be in state $i$, and  $k_{i\to j}$ is the rate of transition from state $i$ to state $j$, with the eight states enumerated as in Figs.~\ref{fig:two-well}B,C. For the detailed-balance-violating case (Fig.~\ref{fig:two-well}B), this gives
\begin{equation} \label{rate-no-DB}
    k_{L\to R}=\frac{\gamma}{4}\,\frac{4\alpha (1+\xi)+\gamma (1+6 \xi+\xi^2)}{2 \alpha+\gamma (1+\xi)} .
\end{equation}
This rate increases with $\alpha$, consistent with the intuition above. It further attains the intuitively appealing limit of $ k_{L\to R} \to  \frac{\gamma}{2} (1+\xi)$  for $\alpha \to \infty$ , where the observed transition rate is simply the average of the slow ($\gamma \xi$) and the fast ($\gamma$) transition rate. 

Contrast this now with the case of Fig.~\ref{fig:two-well}C, where detailed balance is valid; here we obtain from Eq.~\ref{flux-over-pop} 
\begin{equation} \label{rate-DB}
    k_{L\to R}=\frac{2 \gamma \xi}{1+\xi} = \frac{2 \gamma }{1+e^{\beta \epsilon}},
\end{equation}
which is independent of the rate $\alpha$ characterizing the landscape dynamics! In what follows we will show that this independence from the timescale of landscape dynamics remains true for a more general class of models.  

Importantly, while the mean jump rate and, equivalently, the mean waiting time between jumps, 
\begin{equation} \label{mean jump time}
    \tau = \langle t \rangle_L= \langle t \rangle_R \equiv  \int_0^\infty dt\, t\, p_{R,L}(t) = \frac{1}{k_{L\to R}},
\end{equation}
is independent of $\alpha$, {\it the distributions} $p_R(t)$ and $p_L(t)$ of the times that the walker spends on the right and on the left  (here $p_R(t)=p_L(t)$ because of the symmetry of the problem)  do depend on $\alpha$,  as shown in Appendix A.  In particular, in the limit $\alpha \to 0$ we find a bi-exponential distribution,
 \begin{equation} \label{slow limit}
     p_L(t) \to \frac{1}{2}(\gamma e^{-\gamma t}+\gamma \xi e^{-\gamma \xi t}),
 \end{equation}
 which is a linear combination of the distributions for the escape from shallow and deep potential wells. For $\alpha \to \infty$  we find an exponential distribution
\begin{equation}\label{fast limit}
    p_L(t)\to \frac{2\gamma \xi}{1+\xi} e^{-\frac{2\gamma \xi}{1+\xi} t},
\end{equation}
 corresponding to an escape with an effective rate consistent with Eq.~\ref{rate-DB},  
\begin{equation} \label{effective rate}
    k_{L\to R}=\frac{2\gamma \xi}{1+\xi}.
\end{equation}
Notably, this rate is not the arithmetic average of $\gamma$ and $\gamma \xi$ taken with equal weights.

\section{Random walks on a fluctuating lattice}
\begin{figure}
    \centering
    \includegraphics[width=0.75\linewidth]{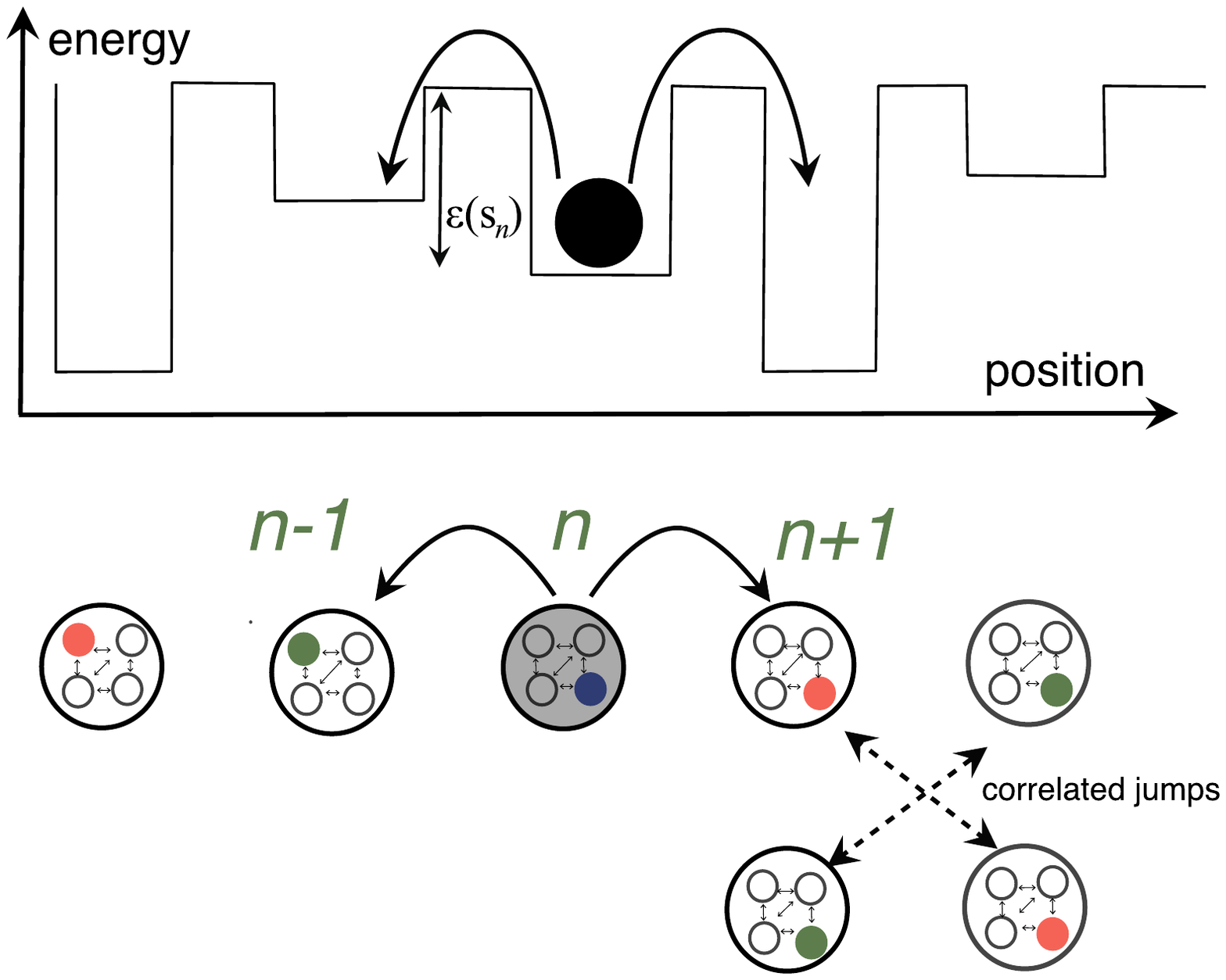}
    \caption{Top: Random walk on a landscape with potential wells of fluctuating depth. Bottom: We model this using a discrete lattice, with each lattice site assuming one of  $S$  possible internal states (here $S=4$). The internal dynamics of the lattice is a Markov jump process, where changes in the internal states of the lattice sites may or may not be correlated. The internal state $s_n$  of a site (and possibly the states of the two adjacent sites, $s_{n\pm 1}$) determines the heights of the potential energy barriers that the walker must overcome when  hopping left or right.}
    \label{fig:lattice-walk}
\end{figure}

Consider now more generally a random walk shown in Fig.~\ref{fig:lattice-walk}. The walker hops among identical lattice sites enumerated by integers $n$. The $n$-th  lattice site can be in one of its $S$ internal states,  which are enumerated by an index $s_n = 1 \dots S$.  The dynamics of interconversion between internal states is a continuous-time Markov jump process. In the simplest possible case, the internal state of each site evolves independently of other sites, as in the model from the previous Section. But we will also consider more complicated lattice dynamics, where the states of different sites are coupled. For example, two or more adjacent sites may change their internal states simultaneously, as illustrated in Fig.~\ref{fig:lattice-walk}.  

When the walker occupies the $n$-th site,  it can hop to the right or left with rates that depend on the internal state $s_n$ of this site and, possibly, on the internal states of the nearby sites. Here, we limit the discussion to the case where these rates depend only on the states of the sites before and after the hop.  Let $k(s_n\to s_{n\pm 1})$ be the rate for jumping from site $n$ to site $n\pm 1$, which generally depends on the internal states $s_n$ and $s_{n+1}$ of the sites. We assume that these rates satisfy the detailed balance condition,
\begin{equation} \label{DB for rates}
    \frac{k(s\to s')}{k(s'\to s)} =e^{-\beta [\epsilon(s)-\epsilon(s')]},
\end{equation}
where the parameter $\epsilon(s)$ can be interpreted as the strength of the walker's binding to the site. We will be especially interested in the case where the jump rates only depend on the internal state of the initial site, 
\begin{equation} \label{rate vs state}
    k(s\to s') \equiv k_{\pm}(s) =k_0 e^{-\beta \epsilon(s)}.
\end{equation}
Eq.~\ref{rate vs state} obviously satisfies Eq.~\ref{DB for rates}.  A physical interpretation  of Eq.~\ref{rate vs state} is that the walker, when located at a site in an internal state $s$, experiences a symmetric potential well whose depth is $\epsilon(s)$, as illustrated in Fig.~\ref{fig:lattice-walk} (top). 

\begin{figure}
    \centering
    \includegraphics[width=0.75\linewidth]{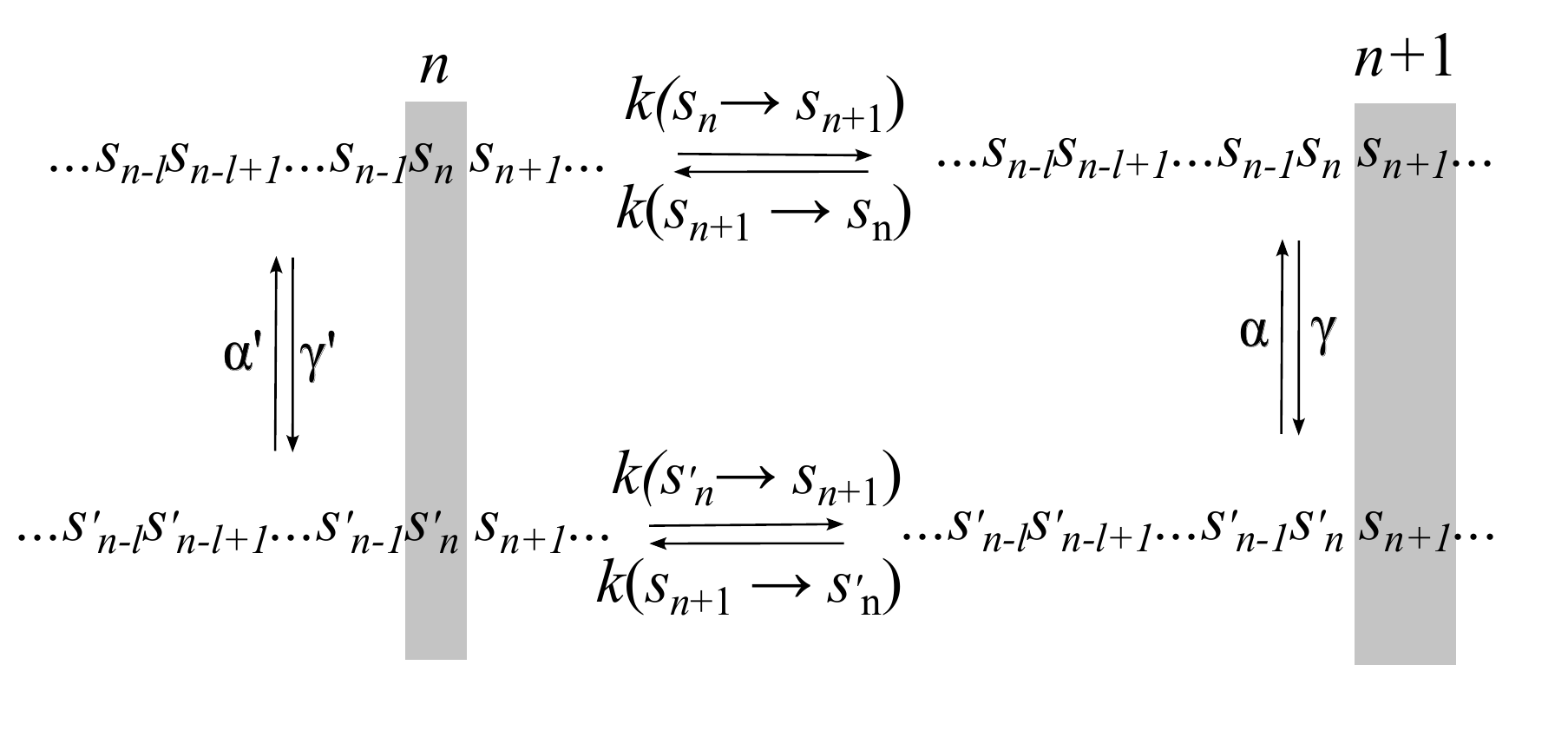}
    \caption{An example of a cycle involving four states of the walker-lattice system. This cycle is part of a larger kinetic network. Transitions in the horizontal direction represent the walker's jumps between sites $n$ and $n+1$, while vertical transitions are between the internal states of the lattice. Here we consider the case  where the internal states of $l+1$ adjacent lattice sites can change simultaneously, $(s_{n-l},\dots,s_n)\leftrightarrow (s'_{n-l},\dots,s'_n) $.  When sites $n-l\dots n$, whose internal states change in a transition,  are empty, the transition rates  are $\alpha$ and $\gamma$. When site $n$ is occupied by the walker, these rates become $\alpha'$ and $\gamma'$, such that detailed balance is satisfied for the cycle.}
    \label{fig:cycle}
\end{figure}

As in the two-well model of the preceding Section, detailed balance necessitates that the presence of the walker alters the rates of the transitions between internal states of the lattice. To understand the consequences of this, consider the four-state cycle shown in Fig.  \ref{fig:cycle}.   In this cycle, horizontal transitions are the walker's hops between sites $n$ and $n+1$, while vertical transitions change internal states of the lattice.
Specifically, to emphasize that our results are not limited to the case where sites are independent, we consider a transition in which the internal states of $l+1$ adjacent sites, $n-l$ to $n$,  change simultaneously. When detailed balance is satisfied, the products of the four rate coefficients for going clockwise and counterclockwise are the same:
\begin{equation} \label{DB-cycle}
    \alpha' k(s_n\to s_{n+ 1})\gamma k(s_{n+1}\to s'_{n})=\alpha k(s_{n+1}\to s_{n})\gamma' k(s'_{n}\to s_{n+1})
\end{equation}
Using  Eqs. \ref{DB for rates} and \ref{DB-cycle},  we have
\begin{equation} \label{DB-cycle1}
    \frac{\alpha'}{\gamma'} \frac{e^{-\beta \epsilon(s_n)}}{e^{-\beta \epsilon(s_{n+1})}}=\frac{\alpha}{\gamma} \frac{e^{-\beta \epsilon(s'_n)}}{e^{-\beta \epsilon(s_{n+1})}}.
\end{equation}
On the other hand, using detailed balance for pairs of vertically connected states, we observe
\begin{equation} \label{prob ratio 1}
   \frac{\alpha'}{\gamma'}= \frac{p(\dots s_{n-l}\dots s_n s_{n+1} \dots  \dots|n)}{p(\dots s'_{n-l} \dots s'_{n} s_{n+1} \dots|n)}
\end{equation}
and 
\begin{equation} \label{prob ratio 2}
    \frac{\alpha}{\gamma} = \frac{p(\dots s_{n-l}\dots s_n s_{n+1} \dots  \dots|n+1)}{p(\dots s'_{n-l} \dots s'_{n} s_{n+1} \dots|n+1)},
\end{equation}
where $p(\vec{s}|n)\equiv p(\dots s_{n-l}\dots s_n s_{n+1} \dots  \dots|n)$ 
is the probability  that the lattice is in the internal state configuration specified by the vector $\vec{s}$  conditional upon the walker occupying site $n$.   But since the rate coefficients $\alpha$ and $\gamma$ are the same as those for the empty lattice, we can also write
\begin{equation} \label{prob ratio 3}
    \frac{\alpha}{\gamma} = \frac{p^0(\dots s_{n-l}\dots s_n s_{n+1} \dots  \dots)}{p^0(\dots s'_{n-l} \dots s'_{n} s_{n+1} \dots)},
\end{equation}
where $p^{0}(\vec{s})\equiv p^{0}(\dots s_{n-1}, s_n, s_{n+1}, \dots)$ is the equilibrium probability of the lattice configuration $\vec{s}$ in the absence of the walker. 
 Equations \ref{DB-cycle1} - \ref{prob ratio 3} are thus satisfied when the coefficients $\alpha'$ and $\gamma'$  are such that the probabilities  $p(\vec{s}|n)$  are proportional to $p^0(\vec{s})e^{\beta \epsilon(s_n)}$. The normalization condition $\sum_{\vec{s}}p(\vec{s}|n)=1$  then leads to
\begin{equation} \label{internal state probability}
    p(\vec{s}|n)=\frac{p^0(\vec{s})e^{\beta \epsilon(s_n)} }{\langle e^{\beta \epsilon} \rangle} ,
\end{equation}
where 
\begin{equation}
    \langle e^{\beta \epsilon} \rangle = \sum_{s=1}^{S}p^0(s)e^{\beta \epsilon(s)}
\end{equation}
and $p^0(s)=\sum_{\vec{s}:s_n=s}p(\vec{s})$ is the marginal equilibrium probability for an empty lattice site to be in the internal state $s$. 

Let us now calculate the mean time $\tau $ that the walker spends on a site. Similarly to Eq.\ref{flux-over-pop}, this is the inverse frequency of jumps, which is equal to the inverse flux leaving the site in both directions: 
\begin{multline} \label{flux}
    J=\tau^{-1}=\sum_{\vec{s}}p(\vec{s}|n)[k(s_n\to s_{n+1})+k(s_n \to s_{n-1})] \\ ={\langle e^{\beta \epsilon} \rangle} ^{-1}\sum_{\vec{s}}p^0(\vec{s})e^{\beta \epsilon(s_n)}[k(s_n\to s_{n+1})+k(s_n \to s_{n-1})]
\end{multline}
Notably, the mean time between the jumps, $\tau$, is unaffected by the dynamics of the lattice and is determined by its {\it equilibrium} statistics. We emphasize that the average denoted by the angular brackets is that over the distribution $p^0(s)$ of a site's internal state for a {\it lattice without the walker}. 

Similarly, the short-time diffusivity, 
\begin{equation} \label{D0}
D(0)=\lim_{t\to 0}\frac{\langle x(t)^2 \rangle}{2t},
\end{equation}
where $x(t$) is the distance traveled by the walker in time $t$, remains independent of the landscape fluctuation dynamics. Indeed, consider the fate of a walker that occupies the $n$-th site after a short interval of time $\delta t$.  It can step right, $n(\delta t)=n+1$, with the probability $\delta t \sum_{\vec{s}}p(\vec{s}|n)k(s_n\to s_{n+1})$,  step left, $n(\delta t)=n-1$, with the probability $\delta t \sum_{\vec{s}}p(\vec{s}|n)k(s_n\to s_{n-1})$,  or stay at $n$  with the probability $1-\delta t \sum_{\vec{s}}p(\vec{s}|n)[k(s_n\to s_{n-1})+k(s_n\to s_{n+1})]$.  The mean square distance traveled is, then,
\begin{equation} \label{msd short time}
    \langle x(\delta t)^2 \rangle=a^2 \langle [n(\delta t)-n]^2 \rangle = a^2 \delta t\sum_{\vec{s}}p(\vec{s}|n)[k(s_n\to s_{n-1})+k(s_n\to s_{n+1})],
\end{equation}
where $a$ is the lattice spacing. Together with Eqs.~\ref{internal state probability} and \ref{D0}, this results in
\begin{equation} \label{short t diff}
    D(0)=\frac{a^2}{2}{\langle e^{\beta \epsilon} \rangle} ^{-1}\sum_{\vec{s}}p^0(\vec{s})e^{\beta \epsilon(s_n)}[k(s_n\to s_{n+1})+k(s_n \to s_{n-1})]=
    \frac{a^2}{2\tau}.
\end{equation}
In the particular case where the jump rates only depend on the departure site, Eq.~\ref{rate vs state}, using Eqs. \ref{flux} and \ref{short t diff} we further find:
\begin{equation} \label {mean waiting time}
    \tau = \frac{1}{2k_0} \langle e^{\beta \epsilon} \rangle, 
\end{equation}
We note that this time is longer than the mean time $(2 k_0)^{-1}$ in the absence of potential wells trapping the walker and, moreover, is longer than the time 
\begin{equation} \label {mean waiting time 1}
    \tau' = \frac{1}{2k_0}  e^{\beta \langle \epsilon \rangle} , 
\end{equation}
which one would observe if all potential wells had the same depth $\langle \epsilon \rangle$. In other words, fluctuations of the energy landscape around its mean decelerate the random walk and reduce the associated diffusivity by a factor $\langle e^{\beta (\epsilon-\langle \epsilon \rangle)} \rangle \ge 1$ .  Also note that we assume that the average  $\langle e^{\beta \epsilon} \rangle$ and, therefore, the time $\tau$ is finite, which is the case when the number of internal states $S$ is finite. Thus phenomena such as aging\cite{Klafter_Sokolov_2016, Barkai2014} do not arise here. 

Since for the model where the hopping rate is independent of the arrival site (Eq.~\ref{rate vs state}), the probabilities of hopping left or right are always equal to $1/2$ for each jump, the mean square displacement of the walker grows strictly linearly with time\footnote{We can write $x(t)=a\sum_{i}\sigma_{i}\theta(t-t_i)$, where $\sigma_i=\pm 1$ are equiprobable Bernoulli random variables, $\theta(t)$ is the Heaviside function and the (ordered) $t_i$ are the jump times. Squaring this and using $\theta^2(t)=\theta(t)$, we find $x(t)^2 = a^2\sum_{i}\theta(t-t_i)+2a^2\sum_{i< j}\sigma_i \sigma_j \theta(t-t_i) \theta(t-t_j)$ .  After taking the average, the second sum disappears because $\sigma_j$ is independent of $\sigma_i$, $t_i$ and $t_j$ for $j>i$. The first sum becomes $a^2 \langle \sum_{i}\theta(t-t_i) \rangle$. Taking the time derivative,  $d\langle x(t)^2 \rangle/dt  = a^2 \langle \sum_i \delta(t-t_i) \rangle = a^2/\tau$, where $\tau$ is the mean time between the jumps.}  
at any $t$,
\begin{equation} \label{msd}
    \langle x(t)^2 \rangle = 2Dt.
\end{equation}
Therefore,  in this case Eq.~\ref{short t diff} gives the  diffusivity not just at short times but at any time,
\begin{equation} \label{diffusivity symmetric}
    D=D(0)=\frac{a^2}{2 \tau}=\frac{k_0 a^2}{\langle e^{\beta \epsilon}\rangle}.
\end{equation}

Not only is the diffusivity in this case independent of  the landscape dynamics, but  it turns out to  be identical to the diffusivity estimate  obtained using Zwanzig's approach based on first passage times  \cite{Zwanzig_1988}, provided that the average indicated by the angular brackets is replaced by the spatial average over a static landscape assumed in ref.\cite{Zwanzig_1988}.  Remarkably, therefore, Lifson-Jackson-Zwanzig's  prediction does not change when the landscape fluctuates in time!

For a more general class of fluctuating energy landscapes, the mean-square displacement is not necessarily a linear function of time\cite{Hughes_2002,Klafter_Sokolov_2016,Demery2011-ak}. Consider, for example, the model with 
\begin{equation} \label{Glauber rate}
    k(s\to s') = k_0 e^{-\beta \lambda \epsilon(s)+\beta (1-\lambda) \epsilon(s')},
\end{equation}
which still satisfies detailed balance, Eq. \ref{DB for rates}, but allows the hopping rates to depend on the state of the arrival sites. Here we assume  $0\le \lambda \le 1$, with $\lambda = 1$ corresponding to Eq. \ref{rate vs state}. 
In this case,  we can still use Eq. \ref{short t diff} to estimate the short-time diffusivity, which gives

\begin{equation} \label{short t diff Glauber}
    D(0)=k_0 a^2\frac{\langle e^{\beta(1-\lambda)[\epsilon(s_n)+\epsilon(s_{n+1})]}\rangle}{\langle e^{\beta \epsilon} \rangle}.
\end{equation}
Again, this diffusivity is determined by the equilibrium properties of the empty lattice and its interaction with the walker  (i.e., by $\epsilon(s)$). We can further consider the following two limits. In the first, the internal states of the neighboring lattice sites are statistically independent.  Eq.~\ref{short t diff Glauber} then becomes  
\begin{equation} \label{short t diff 1}
    D(0)=k_0 a^2 \frac{\langle e^{(1-\lambda)\beta \epsilon }\rangle^2}{\langle e^{\beta \epsilon }\rangle},
\end{equation}
In the second, the characteristic lengthscale of the landscape fluctuations is much longer than the lattice spacing $a$. This case is particularly relevant if one wishes to take a continuous limit of the discrete model. Then we have $\epsilon(s_n)\approx \epsilon(s_{n+1})$ , and 
\begin{equation} \label{short t diff 2}
    D(0)\approx k_0 a^2 \frac{\langle e^{2(1-\lambda)\beta \epsilon }\rangle}{\langle e^{\beta \epsilon }\rangle}.
\end{equation}
Both of these equations become Eq.~\ref{diffusivity symmetric} when $\lambda = 1$. 
At the same time, for $\lambda \ne 1$ both of these results differ from the diffusivity estimated from the Lifson-Jackson-Zwanzig model except in the  limit $\beta \epsilon\ll 1$, where the energy fluctuation is a small perturbation, see Eqs.~\ref{ZD1} and \ref{ZD2}.

In contrast to the lack of simple diffusive behavior in the general case,  Eq.~\ref{flux}  describing the mean time between jumps remains {\it exact}, and so we find
\begin{equation} \label{mean t Glauber 1}
    \tau=\frac{a^2}{2D(0)}=\frac{1}{2k_0}  \frac{\langle e^{\beta \epsilon }\rangle}{\langle e^{(1-\lambda)\beta \epsilon }\rangle^2}
\end{equation}
when the internal states of the lattice are statistically independent, and 
\begin{equation} \label{mean t Glauber 2}
    \tau=\frac{1}{2k_0}  \frac{\langle e^{\beta \epsilon }\rangle}{\langle e^{2(1-\lambda)\beta \epsilon }\rangle}
\end{equation}
in the case of long-wavelength landscape fluctuations.  

To summarize our findings here, first, the short-time diffusivity $D(0)$ and the mean time $\tau$ between jumps for a walker on a fluctuating landscape are independent of the fluctuation dynamics and are entirely determined by the equilibrium statistics of the empty landscape. Here we have proven this explicitly for a model in which the walker's rates of transitions to neighboring sites are determined by the departure and arrival sites, but this can be generalized even for a broader class of models where those rates may depend on other sites. Second, for a model in which the hopping rate only depends on the departure site but not on the arrival site, the dynamics are purely diffusive, with a diffusivity that is independent of the landscape fluctuation timescales. In this case, the mean time between jumps is given by Eq. \ref{mean waiting time} and the short-time diffusivity by Eq. \ref{diffusivity symmetric}, the latter being identical to (a discrete version of) the diffusivity estimate obtained from the Lifson-Jackson-Zwanzig theory assuming a static landscape.  Third, when the jump rates depend  both on the departure and arrival sites, particularly for the model described by Eq. \ref{Glauber rate}, we can obtain further estimates for $\tau$ and for $D(0)$, which, unlike the previous case, will generally depend on {\it spatial correlations between sites}. In other words they cannot be generally determined if only the marginal distribution $p^0(s)$ of a site's internal state is known.    


\section{Concluding remarks}
The discrepancy between diffusivities associated with various dynamical processes observed in biophysical studies and their theoretical estimates are often attributed to the roughness of the underlying free energy landscape, which is not captured by theories (see, e.g., refs.\cite{Hyeon2003,Munoz2007,PhysRevX.14.011017,Khatri2008,Nevo2005,Neupane2017,Hagen2010,Wagner1999,Lapidus2008,Gruebele2009,Nettels2007}). The problem with this interpretation  is that the model of diffusion in a  rough landscape, as previously formulated\cite{Lifson_Jackson_1962,Zwanzig_1988}, does not account for temporal fluctuations of such a landscape.  Naively, such temporal fluctuations are expected to smoothen the landscape and accelerate the observed dynamics -- and indeed, computer simulations and measurements usually yield smooth free energy landscapes, e.g., for the folding of biomolecules\cite{PaiChiTitin,Stirnemann2013,Manuel2015}. Here we show that such an acceleration effect does not necessarily occur, because this naive picture does not account for the back-reaction of the observed system on the landscape. When this back-reaction is taken into account such that the combined system satisfies detailed balance, certain features of the observed dynamics no longer depend on the timescale of the fluctuations of the landscape. As a result, use of the Lifson-Jackson-Zwanzig model of diffusion in a rough landscape may be justifiable even when its  underlying assumptions are invalid.  On the other hand,  except in special cases, landscape fluctuations lead to non-Markovian dynamics, where a single diffusivity parameter cannot explain the dynamics at all timescales, and -- at best -- an {\it effective} diffusivity must be used to describe experimental observations.         
\begin{acknowledgments}
\noindent This work was supported by
 US National Science Foundation (Grant no. CHE 2400424 to DEM),  the Robert A. Welch Foundation (Grant no. F-1514 to  D.E.M.), and the Alexander von Humboldt Foundation. Discussions with Alexander Berezhkovskii, Kristian Blom, Aritra Chowdhury, Aljaz Godec, Gilad Haran, Hagen Hofmann, Matthias Kr\"uger, Benjamin Schuler, and Dave Thirumalai are gratefully acknowledged. 
\end{acknowledgments}

\appendix
\section{Distribution of the lifetime in a well for the model of Fig.\ref{fig:two-well}}
\begin{figure}
    \centering
    \includegraphics[width=0.75\linewidth]{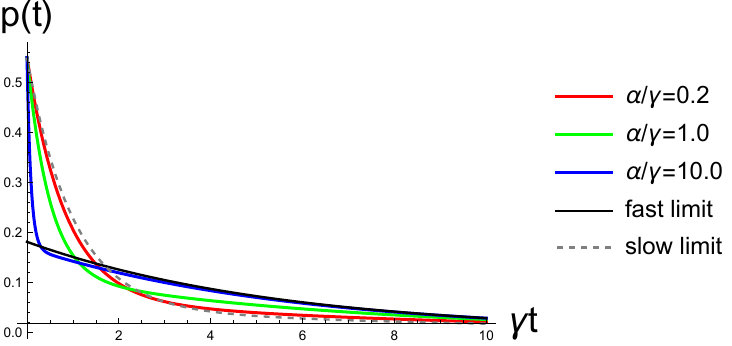}
    \caption{Distribution of the lifetime in the left or right potential well for the model of Fig.~\ref{fig:two-well} plotted for several values of the switching rate $\alpha$. The slow and fast switching limits, Eqs.~\ref{slow limit} and \ref{fast limit}, are also shown. Note that all distributions shown here have identical means.}
    \label{fig:distributions}
\end{figure}

The distribution of the lifetime in the left well,  $p_L(t)$  (we have $p_R(t)=p_L(t)$ because of the symmetry of the problem)  is given by (see, e.g., Ref.\cite{SkinnerPRL})
\begin{equation} \label{Skinner Dunkel}
p_L(t)=-\frac{\mathbf{1}_{L} e^{\mathbf{K}_{L} t}(\mathbf{K}_L)^2 \mathbf{p}_L}{\mathbf{1}_L \mathbf{K}_L \mathbf{p}_L}.
 \end{equation}
 Here 
 \begin{equation}
     \mathbf{K}_L=\begin{bmatrix}
     -2\alpha-\gamma&\alpha \xi& 0& \alpha \\
     \alpha&-\alpha-\alpha \xi-\gamma \xi&\alpha&0 \\
     0&\alpha&-\alpha-\alpha \xi-\gamma \xi&\alpha\\
     \alpha&0&\alpha \xi&-2\alpha-\gamma
     \end{bmatrix}
 \end{equation}
 is a kinetic matrix describing the dynamics of the states 1-4  in Fig.~\ref{fig:two-well} (which, collectively,  correspond to the system being on the left), including diagonal sink terms allowing escape to the right. Furthermore, $\mathbf{p}_L =\frac{1}{4(1+\xi)}(\xi,1,1,\xi)^{\rm T}$  is the vector of equilibrium populations of states 1-4, and $\mathbf{1}_L=(1,1,1,1)^{\rm T}$.  Eq. \ref{Skinner Dunkel} can be evaluated analytically, but the resulting expression is too long to offer further insight, and will not be reproduced here.  The distribution is plotted in Fig. \ref{fig:distributions} for different values of $\alpha$. It can be shown analytically and is also observed in Fig. \ref{fig:distributions} that Eq. \ref{Skinner Dunkel} approaches Eqs. \ref{slow limit} and \ref{fast limit} in the limits $\alpha\to 0$ and $\alpha\to \infty$, respectively. 

\section{The case of quenched disorder: Effective diffusivity from first passage times using Zwanzig's approach.}
Consider a random walk on a one-dimensional lattice of sites enumerated by index $n=0,1,\dots $, with rates $k_{\pm}(n)$ for jumping left and right from a site $n$.
Assume a reflecting boundary at $n=0$ such that $k_-(0)=0$. Consider now the mean first passage time $\tau_{N,M}$ to go to site $N$ starting from site $M>0$. This is given by the following equations \cite{van_Kampen_book}:
\begin{equation} \label{van Kampen1}
    \tau_{N,M}=\sum_{n=M}^{N-1}\Delta_n, 
\end{equation}
\begin{equation} \label{van Kampen2}
    \Delta_n=\tau_{N,n}-\tau_{N,n+1},
\end{equation}
\begin{equation}
     \Delta_0=1/k_+(0),
\end{equation}
\begin{equation} \label{van Kampen3}
    \Delta_n=\frac{1}{k_+(n)} \left(1+\frac{1}{K_n}+\frac{1}{K_n K_{n-1}}+\dots +\frac{1}{K_n K_{n-1}\dots K_1}\right),
\end{equation} 
\begin{equation} \label{van Kampen4}
   K_n=\frac{k_+(n-1)} {k_-(n)}.
\end{equation}
Now suppose that the transition rates satisfy the condition (cf.~Eq.~\ref{DB for rates})
\begin{equation} \label{rate vs state 1}
    K_n=k_{+}(n-1)/k_+(n)=k_0 e^{-\beta (\epsilon_n-\epsilon_{n-1})}, 
\end{equation}
where $\epsilon_n$ characterizes the strength of the walker's interaction with the $n$-th site. Then, using Eqs.~\ref{van Kampen1}--\ref{van Kampen4}, we have 
\begin{equation} \label{VK_MFPT}
    \tau_{N,M}=\sum_{n=M}^{N}\frac{e^{-\beta \epsilon_n}}{k_+(n)}\sum_{n'=0}^{n}e^{\beta \epsilon_{n'}}.
\end{equation}
This is analogous to the mean first passage time from $a$ to $b>a$ in the continuous case\cite{Zwanzig_1988}, 
\begin{equation} \label{continuous 1}
    \tau_{b,a}=\int_{a}^{b}dx \,\frac{e^{\beta U(x)}}{D(x)}\int_{0}^{x}dy \,e^{-\beta U(y)},
\end{equation}
for a particle diffusing in a potential $U(x)$ with a coordinate-dependent diffusivity $D(x)$, where a reflecting boundary  is placed at $x=0<a$.  Note that in our notation $\epsilon_n$ corresponds to the depth of a potential well -- hence the sign difference in the exponent between the two expressions.  If $U(x)$ and $D(x)$ are ``random'' functions with fluctuation lengthscales much shorter than $b-a$, we can follow the approach of  Zwanzig\cite{Zwanzig_1988} and approximate the above expression by 
\begin{equation} \label{continuous 2}
  \tau_{b,a}\approx \int_{a}^{b}dx\, \frac{e^{\beta U(x)}}{D(x)}x \langle e^{-\beta U}\rangle \approx \int_{a}^{b}dx \,\left\langle \frac{e^{\beta U}}{D} \right\rangle x \langle e^{-\beta U}\rangle \approx \int_{a}^{b}dx \frac{1}{D_{\rm eff}}\int_{0}^{x}dy   
\end{equation}
with
\begin{equation} \label{continuous 3}
  \frac{1}{D_{\rm eff}}= \langle e^{\beta U}/D \rangle \langle e^{-\beta U}\rangle , 
\end{equation}
where the angular brackets represent appropriate spatial averages. In particular, if we choose (cf.~Eq.~\ref{rate vs state})   
\begin{equation} \label{rate vs state 1}
    k_{+}(n)=k_{-}(n)=k_0 e^{-\beta \epsilon_n} 
\end{equation}
then 
\begin{equation}
    \tau_{N,M} = k_0^{-1}\sum_{n=M}^{N}\sum_{n'=0}^{n}e^{\beta \epsilon_{n'}} \approx k_0^{-1}\sum_{n=M}^{N}\sum_{n'=0}^{n}\langle e^{\beta \epsilon_{n}} \rangle,
\end{equation}
which is the mean first passage time for a random walk with constant, renormalized jump rates,
\begin{equation}
    k_{\pm}=\frac{k_0}{\langle e^{\beta \epsilon}\rangle} ,
\end{equation}
and thus with renormalized mean waiting time between jumps  
\begin{equation}
    \frac{1}{k_+ +k_-}= \frac{\langle e^{\beta \epsilon} \rangle}{2 k_0},
\end{equation}
and with effective diffusivity  given by 
\begin{equation} \label{ZDsymm}
    D_{\rm eff}=\frac{k_0 a^2}{\langle e^{\beta \epsilon}\rangle}.
\end{equation}
This should be compared to Eq.~\ref{diffusivity symmetric} obtained for our model with dynamically fluctuating disorder. Notice here that the particular choice of $k_{\pm}(n)$ in Eq.~\ref{rate vs state 1} corresponds to $D(x)\propto e^{\beta U(x)}$ in the continuous case. 

Let us now consider a model (cf.~Eq.~\ref{Glauber rate}) where the hopping rates are given by
\begin{equation} \label{Glauber rate static}
    k_{\pm}(n) = k_0 e^{-\beta \lambda \epsilon_n+\beta (1-\lambda) \epsilon_{n\pm 1}}
\end{equation}
with $0\le\lambda\le1$. In this case, we have 
\begin{equation}
\frac{e^{-\beta \epsilon_n}}{k_+(n)}= k_0^{-1}e^{\beta (1-\lambda)[\epsilon_n+\epsilon_{n+1}]}.
\end{equation}
Using Eq.~\ref{VK_MFPT} and following the renormalization procedure of Eqs. \ref{continuous 1}--\ref{continuous 3}, we now obtain an effective hopping rate of the form
\begin{equation}
     k_{\pm}= \frac{k_0}{\langle e^{-\beta (1-\lambda)[\epsilon_n+\epsilon_{n+1}]} \rangle \langle e^{\beta \epsilon_{n}} \rangle},
\end{equation}
or, equivalently, the effective mean time between the hops
\begin{equation}
    \tau = \frac{1}{2 k_0}\langle e^{-\beta (1-\lambda)[\epsilon_n+\epsilon_{n+1}]} \rangle \langle e^{\beta \epsilon_{n}} \rangle.
\end{equation}
Here, the angular brackets should again be interpreted as appropriate spatial averages over the lattice. In particular, if the internal states of neighboring sites are statistically independent then we have
\begin{equation}\label{LJZ 1}
    \tau = \frac{1}{2 k_0}\langle e^{-\beta (1-\lambda)\epsilon} \rangle^2 \langle e^{\beta \epsilon} \rangle, 
\end{equation}
while for long-wavelength correlations, $\epsilon(s_m)\approx \epsilon(s_{m+1})$, we find
\begin{equation} \label{LJZ 2}
    \tau = \frac{1}{2 k_0}\langle e^{-2 \beta (1-\lambda)\epsilon} \rangle \langle e^{\beta \epsilon }\rangle. 
\end{equation}
Following Zwanzig\cite{Zwanzig_1988}, these correspond to diffusivity estimates given by 
\begin{equation}\label{ZD1}
    D_{\rm eff} = \frac{k_0}{\langle e^{-\beta (1-\lambda)\epsilon} \rangle^2 \langle e^{\beta \epsilon} \rangle}, 
\end{equation}
and 
\begin{equation} \label{ZD2}
    D_{\rm eff} = \frac{k_0}{\langle e^{-2 \beta (1-\lambda)\epsilon} \rangle \langle e^{\beta \epsilon }\rangle}. 
\end{equation}
Comparing these with Eqs. \ref{short t diff 1}, \ref{short t diff 2}, \ref{mean t Glauber 1}, and \ref{mean t Glauber 2} we see that they are different except at $\lambda = 1$  and in the perturbative limit $\beta \epsilon \ll 1$. 

Equation \ref{ZD2} agrees with Zwanzig's continuous model\cite{Zwanzig_1988}. Indeed, let $D(x)=D_0 e^{\beta U(x) (2 \lambda -1)}$.  This can be obtained as a continuous limit of Eq. \ref{Glauber rate static} (with $x=an$)  assuming that $\epsilon_n$ is smooth at the lattice scale (which corresponds to the long-wavelength assumption above).  Substituting this into Eq.~\ref{continuous 3}, we get
\begin{equation}
  \frac{1}{D_{\rm eff}}= \frac{1}{D_0} \langle e^{2 \beta U(1-\lambda)} \rangle \langle e^{-\beta U}\rangle , 
\end{equation}
which is the same as Eq.~\ref{ZD2}. 


\bibliography{ms.bib}

\begin{thebibliography}{47}%
\makeatletter
\providecommand \@ifxundefined [1]{%
 \@ifx{#1\undefined}
}%
\providecommand \@ifnum [1]{%
 \ifnum #1\expandafter \@firstoftwo
 \else \expandafter \@secondoftwo
 \fi
}%
\providecommand \@ifx [1]{%
 \ifx #1\expandafter \@firstoftwo
 \else \expandafter \@secondoftwo
 \fi
}%
\providecommand \natexlab [1]{#1}%
\providecommand \enquote  [1]{``#1''}%
\providecommand \bibnamefont  [1]{#1}%
\providecommand \bibfnamefont [1]{#1}%
\providecommand \citenamefont [1]{#1}%
\providecommand \href@noop [0]{\@secondoftwo}%
\providecommand \href [0]{\begingroup \@sanitize@url \@href}%
\providecommand \@href[1]{\@@startlink{#1}\@@href}%
\providecommand \@@href[1]{\endgroup#1\@@endlink}%
\providecommand \@sanitize@url [0]{\catcode `\\12\catcode `\$12\catcode `\&12\catcode `\#12\catcode `\^12\catcode `\_12\catcode `\%12\relax}%
\providecommand \@@startlink[1]{}%
\providecommand \@@endlink[0]{}%
\providecommand \url  [0]{\begingroup\@sanitize@url \@url }%
\providecommand \@url [1]{\endgroup\@href {#1}{\urlprefix }}%
\providecommand \urlprefix  [0]{URL }%
\providecommand \Eprint [0]{\href }%
\providecommand \doibase [0]{http://dx.doi.org/}%
\providecommand \selectlanguage [0]{\@gobble}%
\providecommand \bibinfo  [0]{\@secondoftwo}%
\providecommand \bibfield  [0]{\@secondoftwo}%
\providecommand \translation [1]{[#1]}%
\providecommand \BibitemOpen [0]{}%
\providecommand \bibitemStop [0]{}%
\providecommand \bibitemNoStop [0]{.\EOS\space}%
\providecommand \EOS [0]{\spacefactor3000\relax}%
\providecommand \BibitemShut  [1]{\csname bibitem#1\endcsname}%
\let\auto@bib@innerbib\@empty
\bibitem [{\citenamefont {Makarov}(2015)}]{makarov2015single}%
  \BibitemOpen
  \bibfield  {author} {\bibinfo {author} {\bibfnamefont {D.~E.}\ \bibnamefont {Makarov}},\ }\href@noop {} {\emph {\bibinfo {title} {Single molecule science: physical principles and models}}}\ (\bibinfo  {publisher} {CRC Press},\ \bibinfo {year} {2015})\BibitemShut {NoStop}%
\bibitem [{\citenamefont {Simon}, \citenamefont {Weiss},\ and\ \citenamefont {van Teeffelen}(2024)}]{single_tracking}%
  \BibitemOpen
  \bibfield  {author} {\bibinfo {author} {\bibfnamefont {F.}~\bibnamefont {Simon}}, \bibinfo {author} {\bibfnamefont {L.~E.}\ \bibnamefont {Weiss}}, \ and\ \bibinfo {author} {\bibfnamefont {S.}~\bibnamefont {van Teeffelen}},\ }\bibfield  {title} {\enquote {\bibinfo {title} {A guide to single-particle tracking},}\ }\href {\doibase 10.1038/s43586-024-00341-3} {\bibfield  {journal} {\bibinfo  {journal} {Nature Reviews Methods Primers}\ }\textbf {\bibinfo {volume} {4}},\ \bibinfo {pages} {66} (\bibinfo {year} {2024})}\BibitemShut {NoStop}%
\bibitem [{\citenamefont {Fazel}\ \emph {et~al.}(2024)\citenamefont {Fazel}, \citenamefont {Grussmayer}, \citenamefont {Ferdman}, \citenamefont {Radenovic}, \citenamefont {Shechtman}, \citenamefont {Enderlein},\ and\ \citenamefont {Pressé}}]{Fazel_Grussmayer_Ferdman_Radenovic_Shechtman_Enderlein_Pressé_2024}%
  \BibitemOpen
  \bibfield  {author} {\bibinfo {author} {\bibfnamefont {M.}~\bibnamefont {Fazel}}, \bibinfo {author} {\bibfnamefont {K.~S.}\ \bibnamefont {Grussmayer}}, \bibinfo {author} {\bibfnamefont {B.}~\bibnamefont {Ferdman}}, \bibinfo {author} {\bibfnamefont {A.}~\bibnamefont {Radenovic}}, \bibinfo {author} {\bibfnamefont {Y.}~\bibnamefont {Shechtman}}, \bibinfo {author} {\bibfnamefont {J.}~\bibnamefont {Enderlein}}, \ and\ \bibinfo {author} {\bibfnamefont {S.}~\bibnamefont {Pressé}},\ }\bibfield  {title} {\enquote {\bibinfo {title} {Fluorescence microscopy: A statistics-optics perspective},}\ }\href {\doibase 10.1103/revmodphys.96.025003} {\bibfield  {journal} {\bibinfo  {journal} {Reviews of Modern Physics}\ }\textbf {\bibinfo {volume} {96}} (\bibinfo {year} {2024}),\ 10.1103/revmodphys.96.025003}\BibitemShut {NoStop}%
\bibitem [{\citenamefont {Zwanzig}(2001)}]{Zwanzig}%
  \BibitemOpen
  \bibfield  {author} {\bibinfo {author} {\bibfnamefont {R.}~\bibnamefont {Zwanzig}},\ }\href@noop {} {\emph {\bibinfo {title} {Nonequilibrium Statistical Mechanics}}}\ (\bibinfo  {publisher} {Oxford University Press},\ \bibinfo {year} {2001})\BibitemShut {NoStop}%
\bibitem [{\citenamefont {Mori}(1965)}]{mori1965transport}%
  \BibitemOpen
  \bibfield  {author} {\bibinfo {author} {\bibfnamefont {H.}~\bibnamefont {Mori}},\ }\bibfield  {title} {\enquote {\bibinfo {title} {Transport, collective motion, and brownian motion},}\ }\href@noop {} {\bibfield  {journal} {\bibinfo  {journal} {Progress of theoretical physics}\ }\textbf {\bibinfo {volume} {33}},\ \bibinfo {pages} {423--455} (\bibinfo {year} {1965})}\BibitemShut {NoStop}%
\bibitem [{\citenamefont {Grabert}(2006)}]{grabert2006projection}%
  \BibitemOpen
  \bibfield  {author} {\bibinfo {author} {\bibfnamefont {H.}~\bibnamefont {Grabert}},\ }\href@noop {} {\emph {\bibinfo {title} {Projection operator techniques in nonequilibrium statistical mechanics}}},\ Vol.~\bibinfo {volume} {95}\ (\bibinfo  {publisher} {Springer},\ \bibinfo {year} {2006})\BibitemShut {NoStop}%
\bibitem [{\citenamefont {Glatzel}\ and\ \citenamefont {Schilling}(2021)}]{Glatzel_Schilling_2021}%
  \BibitemOpen
  \bibfield  {author} {\bibinfo {author} {\bibfnamefont {F.}~\bibnamefont {Glatzel}}\ and\ \bibinfo {author} {\bibfnamefont {T.}~\bibnamefont {Schilling}},\ }\bibfield  {title} {\enquote {\bibinfo {title} {The interplay between memory and potentials of mean force: A discussion on the structure of equations of motion for coarse-grained observables},}\ }\href {\doibase 10.1209/0295-5075/ac35ba} {\bibfield  {journal} {\bibinfo  {journal} {Europhysics Letters}\ }\textbf {\bibinfo {volume} {136}},\ \bibinfo {pages} {36001} (\bibinfo {year} {2021})}\BibitemShut {NoStop}%
\bibitem [{\citenamefont {Hij{\'o}n}\ \emph {et~al.}(2010)\citenamefont {Hij{\'o}n}, \citenamefont {Espa{\~n}ol}, \citenamefont {Vanden-Eijnden},\ and\ \citenamefont {Delgado-Buscalioni}}]{hijon2010mori}%
  \BibitemOpen
  \bibfield  {author} {\bibinfo {author} {\bibfnamefont {C.}~\bibnamefont {Hij{\'o}n}}, \bibinfo {author} {\bibfnamefont {P.}~\bibnamefont {Espa{\~n}ol}}, \bibinfo {author} {\bibfnamefont {E.}~\bibnamefont {Vanden-Eijnden}}, \ and\ \bibinfo {author} {\bibfnamefont {R.}~\bibnamefont {Delgado-Buscalioni}},\ }\bibfield  {title} {\enquote {\bibinfo {title} {Mori--zwanzig formalism as a practical computational tool},}\ }\href@noop {} {\bibfield  {journal} {\bibinfo  {journal} {Faraday discussions}\ }\textbf {\bibinfo {volume} {144}},\ \bibinfo {pages} {301--322} (\bibinfo {year} {2010})}\BibitemShut {NoStop}%
\bibitem [{\citenamefont {Ayaz}\ \emph {et~al.}(2022)\citenamefont {Ayaz}, \citenamefont {Scalfi}, \citenamefont {Dalton},\ and\ \citenamefont {Netz}}]{Ayaz2022}%
  \BibitemOpen
  \bibfield  {author} {\bibinfo {author} {\bibfnamefont {C.}~\bibnamefont {Ayaz}}, \bibinfo {author} {\bibfnamefont {L.}~\bibnamefont {Scalfi}}, \bibinfo {author} {\bibfnamefont {B.~A.}\ \bibnamefont {Dalton}}, \ and\ \bibinfo {author} {\bibfnamefont {R.~R.}\ \bibnamefont {Netz}},\ }\bibfield  {title} {\enquote {\bibinfo {title} {Generalized langevin equation with a nonlinear potential of mean force and nonlinear memory friction from a hybrid projection scheme},}\ }\href {\doibase 10.1103/PhysRevE.105.054138} {\bibfield  {journal} {\bibinfo  {journal} {Phys. Rev. E}\ }\textbf {\bibinfo {volume} {105}},\ \bibinfo {pages} {054138} (\bibinfo {year} {2022})}\BibitemShut {NoStop}%
\bibitem [{\citenamefont {Lifson}\ and\ \citenamefont {Jackson}(1962)}]{Lifson_Jackson_1962}%
  \BibitemOpen
  \bibfield  {author} {\bibinfo {author} {\bibfnamefont {S.}~\bibnamefont {Lifson}}\ and\ \bibinfo {author} {\bibfnamefont {J.~L.}\ \bibnamefont {Jackson}},\ }\bibfield  {title} {\enquote {\bibinfo {title} {On the self-diffusion of ions in a polyelectrolyte solution},}\ }\href {\doibase 10.1063/1.1732899} {\bibfield  {journal} {\bibinfo  {journal} {The Journal of Chemical Physics}\ }\textbf {\bibinfo {volume} {36}},\ \bibinfo {pages} {2410–2414} (\bibinfo {year} {1962})}\BibitemShut {NoStop}%
\bibitem [{\citenamefont {Zwanzig}(1988)}]{Zwanzig_1988}%
  \BibitemOpen
  \bibfield  {author} {\bibinfo {author} {\bibfnamefont {R.}~\bibnamefont {Zwanzig}},\ }\bibfield  {title} {\enquote {\bibinfo {title} {Diffusion in a rough potential.}}\ }\href {\doibase 10.1073/pnas.85.7.2029} {\bibfield  {journal} {\bibinfo  {journal} {Proceedings of the National Academy of Sciences}\ }\textbf {\bibinfo {volume} {85}},\ \bibinfo {pages} {2029–2030} (\bibinfo {year} {1988})}\BibitemShut {NoStop}%
\bibitem [{\citenamefont {De~Gennes}(1975)}]{De_Gennes_1975}%
  \BibitemOpen
  \bibfield  {author} {\bibinfo {author} {\bibfnamefont {P.~G.}\ \bibnamefont {De~Gennes}},\ }\bibfield  {title} {\enquote {\bibinfo {title} {Brownian motion of a classical particle through potential barriers. application to the helix-coil transitions of heteropolymers},}\ }\href {\doibase 10.1007/bf01025834} {\bibfield  {journal} {\bibinfo  {journal} {Journal of Statistical Physics}\ }\textbf {\bibinfo {volume} {12}},\ \bibinfo {pages} {463–481} (\bibinfo {year} {1975})}\BibitemShut {NoStop}%
\bibitem [{\citenamefont {Bryngelson}\ and\ \citenamefont {Wolynes}(1989)}]{Bryngelson_Wolynes_1989}%
  \BibitemOpen
  \bibfield  {author} {\bibinfo {author} {\bibfnamefont {J.~D.}\ \bibnamefont {Bryngelson}}\ and\ \bibinfo {author} {\bibfnamefont {P.~G.}\ \bibnamefont {Wolynes}},\ }\bibfield  {title} {\enquote {\bibinfo {title} {Intermediates and barrier crossing in a random energy model (with applications to protein folding)},}\ }\href {\doibase 10.1021/j100356a007} {\bibfield  {journal} {\bibinfo  {journal} {The Journal of Physical Chemistry}\ }\textbf {\bibinfo {volume} {93}},\ \bibinfo {pages} {6902–6915} (\bibinfo {year} {1989})}\BibitemShut {NoStop}%
\bibitem [{\citenamefont {Gray}\ and\ \citenamefont {Yong}(2020)}]{Gray_2020}%
  \BibitemOpen
  \bibfield  {author} {\bibinfo {author} {\bibfnamefont {T.~H.}\ \bibnamefont {Gray}}\ and\ \bibinfo {author} {\bibfnamefont {E.~H.}\ \bibnamefont {Yong}},\ }\bibfield  {title} {\enquote {\bibinfo {title} {Effective diffusion in one-dimensional rough potential-energy landscapes},}\ }\href {\doibase 10.1103/PhysRevE.102.022138} {\bibfield  {journal} {\bibinfo  {journal} {Phys. Rev. E}\ }\textbf {\bibinfo {volume} {102}},\ \bibinfo {pages} {022138} (\bibinfo {year} {2020})}\BibitemShut {NoStop}%
\bibitem [{\citenamefont {Hubbard}\ and\ \citenamefont {Onsager}(1977)}]{Hubbard_Onsager_1977}%
  \BibitemOpen
  \bibfield  {author} {\bibinfo {author} {\bibfnamefont {J.}~\bibnamefont {Hubbard}}\ and\ \bibinfo {author} {\bibfnamefont {L.}~\bibnamefont {Onsager}},\ }\bibfield  {title} {\enquote {\bibinfo {title} {Dielectric dispersion and dielectric friction in electrolyte solutions. i.}}\ }\href {\doibase 10.1063/1.434664} {\bibfield  {journal} {\bibinfo  {journal} {The Journal of Chemical Physics}\ }\textbf {\bibinfo {volume} {67}},\ \bibinfo {pages} {4850–4857} (\bibinfo {year} {1977})}\BibitemShut {NoStop}%
\bibitem [{\citenamefont {Nee}\ and\ \citenamefont {Zwanzig}(1970)}]{Nee_Zwanzig_1970}%
  \BibitemOpen
  \bibfield  {author} {\bibinfo {author} {\bibfnamefont {T.}~\bibnamefont {Nee}}\ and\ \bibinfo {author} {\bibfnamefont {R.}~\bibnamefont {Zwanzig}},\ }\bibfield  {title} {\enquote {\bibinfo {title} {Theory of dielectric relaxation in polar liquids},}\ }\href {\doibase 10.1063/1.1672951} {\bibfield  {journal} {\bibinfo  {journal} {The Journal of Chemical Physics}\ }\textbf {\bibinfo {volume} {52}},\ \bibinfo {pages} {6353--6363} (\bibinfo {year} {1970})}\BibitemShut {NoStop}%
\bibitem [{\citenamefont {Samanta}\ and\ \citenamefont {Matyushov}(2021)}]{Samanta_Matyushov_2021}%
  \BibitemOpen
  \bibfield  {author} {\bibinfo {author} {\bibfnamefont {T.}~\bibnamefont {Samanta}}\ and\ \bibinfo {author} {\bibfnamefont {D.~V.}\ \bibnamefont {Matyushov}},\ }\bibfield  {title} {\enquote {\bibinfo {title} {Dielectric friction, violation of the stokes-einstein-debye relation, and non-gaussian transport dynamics of dipolar solutes in water},}\ }\href {\doibase 10.1103/physrevresearch.3.023025} {\bibfield  {journal} {\bibinfo  {journal} {Physical Review Research}\ }\textbf {\bibinfo {volume} {3}} (\bibinfo {year} {2021}),\ 10.1103/physrevresearch.3.023025}\BibitemShut {NoStop}%
\bibitem [{\citenamefont {Makarov}\ and\ \citenamefont {Hofmann}(2021)}]{Hagen_electric_friction}%
  \BibitemOpen
  \bibfield  {author} {\bibinfo {author} {\bibfnamefont {D.~E.}\ \bibnamefont {Makarov}}\ and\ \bibinfo {author} {\bibfnamefont {H.}~\bibnamefont {Hofmann}},\ }\bibfield  {title} {\enquote {\bibinfo {title} {Does electric friction matter in living cells?}}\ }\href {\doibase 10.1021/acs.jpcb.1c02783} {\bibfield  {journal} {\bibinfo  {journal} {The Journal of Physical Chemistry B}\ }\textbf {\bibinfo {volume} {125}},\ \bibinfo {pages} {6144--6153} (\bibinfo {year} {2021})}\BibitemShut {NoStop}%
\bibitem [{\citenamefont {Straub}, \citenamefont {Borkovec},\ and\ \citenamefont {Berne}(1986)}]{Straub1986}%
  \BibitemOpen
  \bibfield  {author} {\bibinfo {author} {\bibfnamefont {J.}~\bibnamefont {Straub}}, \bibinfo {author} {\bibfnamefont {M.}~\bibnamefont {Borkovec}}, \ and\ \bibinfo {author} {\bibfnamefont {B.}~\bibnamefont {Berne}},\ }\bibfield  {title} {\enquote {\bibinfo {title} {Non-markovian activated rate processes: Comparison of current theories with numerical simulation data},}\ }\href@noop {} {\bibfield  {journal} {\bibinfo  {journal} {J Chem Phys}\ }\textbf {\bibinfo {volume} {84}},\ \bibinfo {pages} {1788--1794} (\bibinfo {year} {1986})}\BibitemShut {NoStop}%
\bibitem [{\citenamefont {Straub}, \citenamefont {Borkovec},\ and\ \citenamefont {Berne}(1987)}]{Straub_1987}%
  \BibitemOpen
  \bibfield  {author} {\bibinfo {author} {\bibfnamefont {J.~E.}\ \bibnamefont {Straub}}, \bibinfo {author} {\bibfnamefont {M.}~\bibnamefont {Borkovec}}, \ and\ \bibinfo {author} {\bibfnamefont {B.~J.}\ \bibnamefont {Berne}},\ }\bibfield  {title} {\enquote {\bibinfo {title} {Calculation of dynamic friction on intramolecular degrees of freedom},}\ }\href@noop {} {\bibfield  {journal} {\bibinfo  {journal} {J Phys Chem}\ }\textbf {\bibinfo {volume} {91}},\ \bibinfo {pages} {4995--4998} (\bibinfo {year} {1987})}\BibitemShut {NoStop}%
\bibitem [{\citenamefont {Kiefer}, \citenamefont {Dalton},\ and\ \citenamefont {Netz}(2025)}]{Netz2025}%
  \BibitemOpen
  \bibfield  {author} {\bibinfo {author} {\bibfnamefont {H.}~\bibnamefont {Kiefer}}, \bibinfo {author} {\bibfnamefont {B.~A.}\ \bibnamefont {Dalton}}, \ and\ \bibinfo {author} {\bibfnamefont {R.~R.}\ \bibnamefont {Netz}},\ }\bibfield  {title} {\enquote {\bibinfo {title} {Diffusion and friction from force correlations},}\ }\href {\doibase 10.1063/5.0261012} {\bibfield  {journal} {\bibinfo  {journal} {The Journal of Chemical Physics}\ }\textbf {\bibinfo {volume} {162}} (\bibinfo {year} {2025}),\ 10.1063/5.0261012}\BibitemShut {NoStop}%
\bibitem [{\citenamefont {Makarov}(2013)}]{MakarovLangevin2013}%
  \BibitemOpen
  \bibfield  {author} {\bibinfo {author} {\bibfnamefont {D.~E.}\ \bibnamefont {Makarov}},\ }\bibfield  {title} {\enquote {\bibinfo {title} {Interplay of non-markov and internal friction effects in the barrier crossing kinetics of unfolded proteins},}\ }\href@noop {} {\bibfield  {journal} {\bibinfo  {journal} {J. Chem. Phys.}\ }\textbf {\bibinfo {volume} {138}},\ \bibinfo {pages} {014102} (\bibinfo {year} {2013})}\BibitemShut {NoStop}%
\bibitem [{\citenamefont {Elber}, \citenamefont {Makarov},\ and\ \citenamefont {Orland}(2020)}]{elber2020molecular}%
  \BibitemOpen
  \bibfield  {author} {\bibinfo {author} {\bibfnamefont {R.}~\bibnamefont {Elber}}, \bibinfo {author} {\bibfnamefont {D.~E.}\ \bibnamefont {Makarov}}, \ and\ \bibinfo {author} {\bibfnamefont {H.}~\bibnamefont {Orland}},\ }\href@noop {} {\emph {\bibinfo {title} {Molecular kinetics in condensed phases: theory, simulation, and analysis}}}\ (\bibinfo  {publisher} {John Wiley \& Sons},\ \bibinfo {year} {2020})\BibitemShut {NoStop}%
\bibitem [{\citenamefont {Mostajabi~Sarhangi}\ and\ \citenamefont {Matyushov}(2024)}]{Matyushov_2024}%
  \BibitemOpen
  \bibfield  {author} {\bibinfo {author} {\bibfnamefont {S.}~\bibnamefont {Mostajabi~Sarhangi}}\ and\ \bibinfo {author} {\bibfnamefont {D.~V.}\ \bibnamefont {Matyushov}},\ }\bibfield  {title} {\enquote {\bibinfo {title} {Remarkable insensitivity of protein diffusion to protein charge},}\ }\href {\doibase 10.1021/acs.jpclett.4c02062} {\bibfield  {journal} {\bibinfo  {journal} {The Journal of Physical Chemistry Letters}\ }\textbf {\bibinfo {volume} {15}},\ \bibinfo {pages} {9502--9508} (\bibinfo {year} {2024})}\BibitemShut {NoStop}%
\bibitem [{\citenamefont {Dean}, \citenamefont {Drummond},\ and\ \citenamefont {Horgan}(2007)}]{Dean2007-bq}%
  \BibitemOpen
  \bibfield  {author} {\bibinfo {author} {\bibfnamefont {D.~S.}\ \bibnamefont {Dean}}, \bibinfo {author} {\bibfnamefont {I.~T.}\ \bibnamefont {Drummond}}, \ and\ \bibinfo {author} {\bibfnamefont {R.~R.}\ \bibnamefont {Horgan}},\ }\bibfield  {title} {\enquote {\bibinfo {title} {Effective transport properties for diffusion in random media},}\ }\href@noop {} {\bibfield  {journal} {\bibinfo  {journal} {J. Stat. Mech.}\ }\textbf {\bibinfo {volume} {2007}},\ \bibinfo {pages} {P07013--P07013} (\bibinfo {year} {2007})}\BibitemShut {NoStop}%
\bibitem [{\citenamefont {Dean}\ and\ \citenamefont {D{\'e}mery}(2011)}]{Dean2011-up}%
  \BibitemOpen
  \bibfield  {author} {\bibinfo {author} {\bibfnamefont {D.~S.}\ \bibnamefont {Dean}}\ and\ \bibinfo {author} {\bibfnamefont {V.}~\bibnamefont {D{\'e}mery}},\ }\bibfield  {title} {\enquote {\bibinfo {title} {Diffusion of active tracers in fluctuating fields},}\ }\href@noop {} {\bibfield  {journal} {\bibinfo  {journal} {J. Phys. Condens. Matter}\ }\textbf {\bibinfo {volume} {23}},\ \bibinfo {pages} {234114} (\bibinfo {year} {2011})}\BibitemShut {NoStop}%
\bibitem [{\citenamefont {D{\'e}mery}\ and\ \citenamefont {Dean}(2011)}]{Demery2011-ak}%
  \BibitemOpen
  \bibfield  {author} {\bibinfo {author} {\bibfnamefont {V.}~\bibnamefont {D{\'e}mery}}\ and\ \bibinfo {author} {\bibfnamefont {D.~S.}\ \bibnamefont {Dean}},\ }\bibfield  {title} {\enquote {\bibinfo {title} {Perturbative path-integral study of active- and passive-tracer diffusion in fluctuating fields},}\ }\href@noop {} {\bibfield  {journal} {\bibinfo  {journal} {Phys. Rev. E Stat. Nonlin. Soft Matter Phys.}\ }\textbf {\bibinfo {volume} {84}},\ \bibinfo {pages} {011148} (\bibinfo {year} {2011})}\BibitemShut {NoStop}%
\bibitem [{\citenamefont {Klafter}\ and\ \citenamefont {Sokolov}(2016)}]{Klafter_Sokolov_2016}%
  \BibitemOpen
  \bibfield  {author} {\bibinfo {author} {\bibfnamefont {J.}~\bibnamefont {Klafter}}\ and\ \bibinfo {author} {\bibfnamefont {I.~M.}\ \bibnamefont {Sokolov}},\ }\href@noop {} {\emph {\bibinfo {title} {First steps in random walks: From tools to applications}}}\ (\bibinfo  {publisher} {Oxford University Press, USA},\ \bibinfo {year} {2016})\BibitemShut {NoStop}%
\bibitem [{\citenamefont {Metzler}\ \emph {et~al.}(2014)\citenamefont {Metzler}, \citenamefont {Jeon}, \citenamefont {Cherstvy},\ and\ \citenamefont {Barkai}}]{Barkai2014}%
  \BibitemOpen
  \bibfield  {author} {\bibinfo {author} {\bibfnamefont {R.}~\bibnamefont {Metzler}}, \bibinfo {author} {\bibfnamefont {J.~H.}\ \bibnamefont {Jeon}}, \bibinfo {author} {\bibfnamefont {A.~G.}\ \bibnamefont {Cherstvy}}, \ and\ \bibinfo {author} {\bibfnamefont {E.}~\bibnamefont {Barkai}},\ }\bibfield  {title} {\enquote {\bibinfo {title} {Anomalous diffusion models and their properties: non-stationarity, non-ergodicity, and ageing at the centenary of single particle tracking},}\ }\href {\doibase 10.1039/c4cp03465a} {\bibfield  {journal} {\bibinfo  {journal} {Phys Chem Chem Phys}\ }\textbf {\bibinfo {volume} {16}},\ \bibinfo {pages} {24128--64} (\bibinfo {year} {2014})}\BibitemShut {NoStop}%
\bibitem [{Note1()}]{Note1}%
  \BibitemOpen
  \bibinfo {note} {We can write $x(t)=a\DOTSB \sum@ \slimits@ _{i}\sigma _{i}\theta (t-t_i)$, where $\sigma _i=\pm 1$ are equiprobable Bernoulli random variables, $\theta (t)$ is the Heaviside function and the (ordered) $t_i$ are the jump times. Squaring this and using $\theta ^2(t)=\theta (t)$, we find $x(t)^2 = a^2\DOTSB \sum@ \slimits@ _{i}\theta (t-t_i)+2a^2\DOTSB \sum@ \slimits@ _{i< j}\sigma _i \sigma _j \theta (t-t_i) \theta (t-t_j)$ . After taking the average, the second sum disappears because $\sigma _j$ is independent of $\sigma _i$, $t_i$ and $t_j$ for $j>i$. The first sum becomes $a^2 \langle \DOTSB \sum@ \slimits@ _{i}\theta (t-t_i) \rangle $. Taking the time derivative, $d\langle x(t)^2 \rangle /dt = a^2 \langle \DOTSB \sum@ \slimits@ _i \delta (t-t_i) \rangle = a^2/\tau $, where $\tau $ is the mean time between the jumps.}\BibitemShut {Stop}%
\bibitem [{\citenamefont {Hughes}(2002)}]{Hughes_2002}%
  \BibitemOpen
  \bibfield  {author} {\bibinfo {author} {\bibfnamefont {B.~D.}\ \bibnamefont {Hughes}},\ }\href@noop {} {\emph {\bibinfo {title} {Random walks and random environments}}}\ (\bibinfo  {publisher} {Clarendon Press},\ \bibinfo {year} {2002})\BibitemShut {NoStop}%
\bibitem [{\citenamefont {Hyeon}\ and\ \citenamefont {Thirumalai}(2003)}]{Hyeon2003}%
  \BibitemOpen
  \bibfield  {author} {\bibinfo {author} {\bibfnamefont {C.}~\bibnamefont {Hyeon}}\ and\ \bibinfo {author} {\bibfnamefont {D.}~\bibnamefont {Thirumalai}},\ }\bibfield  {title} {\enquote {\bibinfo {title} {Can energy landscape roughness of proteins and rna be measured by using mechanical unfolding experiments?}}\ }\href {\doibase 10.1073/pnas.1833310100} {\bibfield  {journal} {\bibinfo  {journal} {Proceedings of the National Academy of Sciences}\ }\textbf {\bibinfo {volume} {100}},\ \bibinfo {pages} {10249--10253} (\bibinfo {year} {2003})},\ \Eprint {http://arxiv.org/abs/https://www.pnas.org/doi/pdf/10.1073/pnas.1833310100} {https://www.pnas.org/doi/pdf/10.1073/pnas.1833310100} \BibitemShut {NoStop}%
\bibitem [{\citenamefont {Naganathan}, \citenamefont {Doshi},\ and\ \citenamefont {Muñoz}(2007)}]{Munoz2007}%
  \BibitemOpen
  \bibfield  {author} {\bibinfo {author} {\bibfnamefont {A.~N.}\ \bibnamefont {Naganathan}}, \bibinfo {author} {\bibfnamefont {U.}~\bibnamefont {Doshi}}, \ and\ \bibinfo {author} {\bibfnamefont {V.}~\bibnamefont {Muñoz}},\ }\bibfield  {title} {\enquote {\bibinfo {title} {Protein folding kinetics: barrier effects in chemical and thermal denaturation experiments},}\ }\href {\doibase 10.1021/ja0689740} {\bibfield  {journal} {\bibinfo  {journal} {J Am Chem Soc}\ }\textbf {\bibinfo {volume} {129}},\ \bibinfo {pages} {5673--82} (\bibinfo {year} {2007})}\BibitemShut {NoStop}%
\bibitem [{\citenamefont {Lyons}\ \emph {et~al.}(2024)\citenamefont {Lyons}, \citenamefont {Devi}, \citenamefont {Hoffer},\ and\ \citenamefont {Woodside}}]{PhysRevX.14.011017}%
  \BibitemOpen
  \bibfield  {author} {\bibinfo {author} {\bibfnamefont {A.}~\bibnamefont {Lyons}}, \bibinfo {author} {\bibfnamefont {A.}~\bibnamefont {Devi}}, \bibinfo {author} {\bibfnamefont {N.~Q.}\ \bibnamefont {Hoffer}}, \ and\ \bibinfo {author} {\bibfnamefont {M.~T.}\ \bibnamefont {Woodside}},\ }\bibfield  {title} {\enquote {\bibinfo {title} {Quantifying the properties of nonproductive attempts at thermally activated energy-barrier crossing through direct observation},}\ }\href {\doibase 10.1103/PhysRevX.14.011017} {\bibfield  {journal} {\bibinfo  {journal} {Phys. Rev. X}\ }\textbf {\bibinfo {volume} {14}},\ \bibinfo {pages} {011017} (\bibinfo {year} {2024})}\BibitemShut {NoStop}%
\bibitem [{\citenamefont {Khatri}\ \emph {et~al.}(2008)\citenamefont {Khatri}, \citenamefont {Byrne}, \citenamefont {Kawakami}, \citenamefont {Brockwell}, \citenamefont {Smith}, \citenamefont {Radford},\ and\ \citenamefont {McLeish}}]{Khatri2008}%
  \BibitemOpen
  \bibfield  {author} {\bibinfo {author} {\bibfnamefont {B.~S.}\ \bibnamefont {Khatri}}, \bibinfo {author} {\bibfnamefont {K.}~\bibnamefont {Byrne}}, \bibinfo {author} {\bibfnamefont {M.}~\bibnamefont {Kawakami}}, \bibinfo {author} {\bibfnamefont {D.~J.}\ \bibnamefont {Brockwell}}, \bibinfo {author} {\bibfnamefont {D.~A.}\ \bibnamefont {Smith}}, \bibinfo {author} {\bibfnamefont {S.~E.}\ \bibnamefont {Radford}}, \ and\ \bibinfo {author} {\bibfnamefont {T.~C.~B.}\ \bibnamefont {McLeish}},\ }\bibfield  {title} {\enquote {\bibinfo {title} {Internal friction of single polypeptide chains at high stretch},}\ }\href {\doibase 10.1039/B716418C} {\bibfield  {journal} {\bibinfo  {journal} {Faraday Discuss.}\ }\textbf {\bibinfo {volume} {139}},\ \bibinfo {pages} {35--51} (\bibinfo {year} {2008})}\BibitemShut {NoStop}%
\bibitem [{\citenamefont {Nevo}\ \emph {et~al.}(2005)\citenamefont {Nevo}, \citenamefont {Brumfeld}, \citenamefont {Kapon}, \citenamefont {Hinterdorfer},\ and\ \citenamefont {Reich}}]{Nevo2005}%
  \BibitemOpen
  \bibfield  {author} {\bibinfo {author} {\bibfnamefont {R.}~\bibnamefont {Nevo}}, \bibinfo {author} {\bibfnamefont {V.}~\bibnamefont {Brumfeld}}, \bibinfo {author} {\bibfnamefont {R.}~\bibnamefont {Kapon}}, \bibinfo {author} {\bibfnamefont {P.}~\bibnamefont {Hinterdorfer}}, \ and\ \bibinfo {author} {\bibfnamefont {Z.}~\bibnamefont {Reich}},\ }\bibfield  {title} {\enquote {\bibinfo {title} {Direct measurement of protein energy landscape roughness},}\ }\href {\doibase 10.1038/sj.embor.7400403} {\bibfield  {journal} {\bibinfo  {journal} {EMBO Rep}\ }\textbf {\bibinfo {volume} {6}},\ \bibinfo {pages} {482--6} (\bibinfo {year} {2005})}\BibitemShut {NoStop}%
\bibitem [{\citenamefont {Neupane}, \citenamefont {Wang},\ and\ \citenamefont {Woodside}(2017)}]{Neupane2017}%
  \BibitemOpen
  \bibfield  {author} {\bibinfo {author} {\bibfnamefont {K.}~\bibnamefont {Neupane}}, \bibinfo {author} {\bibfnamefont {F.}~\bibnamefont {Wang}}, \ and\ \bibinfo {author} {\bibfnamefont {M.~T.}\ \bibnamefont {Woodside}},\ }\bibfield  {title} {\enquote {\bibinfo {title} {Direct measurement of sequence-dependent transition path times and conformational diffusion in dna duplex formation},}\ }\href {\doibase 10.1073/pnas.1611602114} {\bibfield  {journal} {\bibinfo  {journal} {Proc Natl Acad Sci U S A}\ }\textbf {\bibinfo {volume} {114}},\ \bibinfo {pages} {1329--1334} (\bibinfo {year} {2017})}\BibitemShut {NoStop}%
\bibitem [{\citenamefont {Hofmann}\ \emph {et~al.}(2010)\citenamefont {Hofmann}, \citenamefont {Hillger}, \citenamefont {Pfeil}, \citenamefont {Hoffmann}, \citenamefont {Streich}, \citenamefont {Haenni}, \citenamefont {Nettels}, \citenamefont {Lipman},\ and\ \citenamefont {Schuler}}]{Hagen2010}%
  \BibitemOpen
  \bibfield  {author} {\bibinfo {author} {\bibfnamefont {H.}~\bibnamefont {Hofmann}}, \bibinfo {author} {\bibfnamefont {F.}~\bibnamefont {Hillger}}, \bibinfo {author} {\bibfnamefont {S.~H.}\ \bibnamefont {Pfeil}}, \bibinfo {author} {\bibfnamefont {A.}~\bibnamefont {Hoffmann}}, \bibinfo {author} {\bibfnamefont {D.}~\bibnamefont {Streich}}, \bibinfo {author} {\bibfnamefont {D.}~\bibnamefont {Haenni}}, \bibinfo {author} {\bibfnamefont {D.}~\bibnamefont {Nettels}}, \bibinfo {author} {\bibfnamefont {E.~A.}\ \bibnamefont {Lipman}}, \ and\ \bibinfo {author} {\bibfnamefont {B.}~\bibnamefont {Schuler}},\ }\bibfield  {title} {\enquote {\bibinfo {title} {Single-molecule spectroscopy of protein folding in a chaperonin cage},}\ }\href {\doibase 10.1073/pnas.1002356107} {\bibfield  {journal} {\bibinfo  {journal} {Proc Natl Acad Sci U S A}\ }\textbf {\bibinfo {volume} {107}},\ \bibinfo {pages} {11793--8} (\bibinfo {year} {2010})}\BibitemShut {NoStop}%
\bibitem [{\citenamefont {Wagner}\ and\ \citenamefont {Kiefhaber}(1999)}]{Wagner1999}%
  \BibitemOpen
  \bibfield  {author} {\bibinfo {author} {\bibfnamefont {C.}~\bibnamefont {Wagner}}\ and\ \bibinfo {author} {\bibfnamefont {T.}~\bibnamefont {Kiefhaber}},\ }\bibfield  {title} {\enquote {\bibinfo {title} {Intermediates can accelerate protein folding},}\ }\href {\doibase 10.1073/pnas.96.12.6716} {\bibfield  {journal} {\bibinfo  {journal} {Proc Natl Acad Sci U S A}\ }\textbf {\bibinfo {volume} {96}},\ \bibinfo {pages} {6716--21} (\bibinfo {year} {1999})}\BibitemShut {NoStop}%
\bibitem [{\citenamefont {Waldauer}\ \emph {et~al.}(2008)\citenamefont {Waldauer}, \citenamefont {Bakajin}, \citenamefont {Ball}, \citenamefont {Chen}, \citenamefont {Decamp}, \citenamefont {Kopka}, \citenamefont {Jäger}, \citenamefont {Singh}, \citenamefont {Wedemeyer}, \citenamefont {Weiss}, \citenamefont {Yao},\ and\ \citenamefont {Lapidus}}]{Lapidus2008}%
  \BibitemOpen
  \bibfield  {author} {\bibinfo {author} {\bibfnamefont {S.~A.}\ \bibnamefont {Waldauer}}, \bibinfo {author} {\bibfnamefont {O.}~\bibnamefont {Bakajin}}, \bibinfo {author} {\bibfnamefont {T.}~\bibnamefont {Ball}}, \bibinfo {author} {\bibfnamefont {Y.}~\bibnamefont {Chen}}, \bibinfo {author} {\bibfnamefont {S.~J.}\ \bibnamefont {Decamp}}, \bibinfo {author} {\bibfnamefont {M.}~\bibnamefont {Kopka}}, \bibinfo {author} {\bibfnamefont {M.}~\bibnamefont {Jäger}}, \bibinfo {author} {\bibfnamefont {V.~R.}\ \bibnamefont {Singh}}, \bibinfo {author} {\bibfnamefont {W.~J.}\ \bibnamefont {Wedemeyer}}, \bibinfo {author} {\bibfnamefont {S.}~\bibnamefont {Weiss}}, \bibinfo {author} {\bibfnamefont {S.}~\bibnamefont {Yao}}, \ and\ \bibinfo {author} {\bibfnamefont {L.~J.}\ \bibnamefont {Lapidus}},\ }\bibfield  {title} {\enquote {\bibinfo {title} {Ruggedness in the folding landscape of protein l},}\ }\href {\doibase 10.2976/1.3013702} {\bibfield  {journal} {\bibinfo  {journal} {Hfsp j}\ }\textbf {\bibinfo {volume} {2}},\
  \bibinfo {pages} {388--95} (\bibinfo {year} {2008})}\BibitemShut {NoStop}%
\bibitem [{\citenamefont {Liu}, \citenamefont {Nakaema},\ and\ \citenamefont {Gruebele}(2009)}]{Gruebele2009}%
  \BibitemOpen
  \bibfield  {author} {\bibinfo {author} {\bibfnamefont {F.}~\bibnamefont {Liu}}, \bibinfo {author} {\bibfnamefont {M.}~\bibnamefont {Nakaema}}, \ and\ \bibinfo {author} {\bibfnamefont {M.}~\bibnamefont {Gruebele}},\ }\bibfield  {title} {\enquote {\bibinfo {title} {The transition state transit time of ww domain folding is controlled by energy landscape roughness},}\ }\href {\doibase 10.1063/1.3262489} {\bibfield  {journal} {\bibinfo  {journal} {The Journal of Chemical Physics}\ }\textbf {\bibinfo {volume} {131}} (\bibinfo {year} {2009}),\ 10.1063/1.3262489}\BibitemShut {NoStop}%
\bibitem [{\citenamefont {Nettels}\ \emph {et~al.}(2007)\citenamefont {Nettels}, \citenamefont {Gopich}, \citenamefont {Hoffmann},\ and\ \citenamefont {Schuler}}]{Nettels2007}%
  \BibitemOpen
  \bibfield  {author} {\bibinfo {author} {\bibfnamefont {D.}~\bibnamefont {Nettels}}, \bibinfo {author} {\bibfnamefont {I.~V.}\ \bibnamefont {Gopich}}, \bibinfo {author} {\bibfnamefont {A.}~\bibnamefont {Hoffmann}}, \ and\ \bibinfo {author} {\bibfnamefont {B.}~\bibnamefont {Schuler}},\ }\bibfield  {title} {\enquote {\bibinfo {title} {Ultrafast dynamics of protein collapse from single-molecule photon statistics},}\ }\href {\doibase 10.1073/pnas.0611093104} {\bibfield  {journal} {\bibinfo  {journal} {Proc Natl Acad Sci U S A}\ }\textbf {\bibinfo {volume} {104}},\ \bibinfo {pages} {2655--60} (\bibinfo {year} {2007})}\BibitemShut {NoStop}%
\bibitem [{\citenamefont {Li}\ and\ \citenamefont {Makarov}(2003)}]{PaiChiTitin}%
  \BibitemOpen
  \bibfield  {author} {\bibinfo {author} {\bibfnamefont {P.-C.}\ \bibnamefont {Li}}\ and\ \bibinfo {author} {\bibfnamefont {D.~E.}\ \bibnamefont {Makarov}},\ }\bibfield  {title} {\enquote {\bibinfo {title} {Theoretical studies of the mechanical unfolding of the muscle protein titin: Bridging the time-scale gap between simulation and experiment.}}\ }\href@noop {} {\bibfield  {journal} {\bibinfo  {journal} {J. Chem. Phys.}\ }\textbf {\bibinfo {volume} {119}},\ \bibinfo {pages} {9260} (\bibinfo {year} {2003})}\BibitemShut {NoStop}%
\bibitem [{\citenamefont {Stirnemann}\ \emph {et~al.}(2013)\citenamefont {Stirnemann}, \citenamefont {Giganti}, \citenamefont {Fernandez},\ and\ \citenamefont {Berne}}]{Stirnemann2013}%
  \BibitemOpen
  \bibfield  {author} {\bibinfo {author} {\bibfnamefont {G.}~\bibnamefont {Stirnemann}}, \bibinfo {author} {\bibfnamefont {D.}~\bibnamefont {Giganti}}, \bibinfo {author} {\bibfnamefont {J.~M.}\ \bibnamefont {Fernandez}}, \ and\ \bibinfo {author} {\bibfnamefont {B.~J.}\ \bibnamefont {Berne}},\ }\bibfield  {title} {\enquote {\bibinfo {title} {Elasticity, structure, and relaxation of extended proteins under force},}\ }\href {\doibase 10.1073/pnas.1300596110} {\bibfield  {journal} {\bibinfo  {journal} {Proc Natl Acad Sci U S A}\ }\textbf {\bibinfo {volume} {110}},\ \bibinfo {pages} {3847--52} (\bibinfo {year} {2013})}\BibitemShut {NoStop}%
\bibitem [{\citenamefont {Manuel}, \citenamefont {Lambert},\ and\ \citenamefont {Woodside}(2015)}]{Manuel2015}%
  \BibitemOpen
  \bibfield  {author} {\bibinfo {author} {\bibfnamefont {A.~P.}\ \bibnamefont {Manuel}}, \bibinfo {author} {\bibfnamefont {J.}~\bibnamefont {Lambert}}, \ and\ \bibinfo {author} {\bibfnamefont {M.~T.}\ \bibnamefont {Woodside}},\ }\bibfield  {title} {\enquote {\bibinfo {title} {Reconstructing folding energy landscapes from splitting probability analysis of single-molecule trajectories},}\ }\href {\doibase 10.1073/pnas.1419490112} {\bibfield  {journal} {\bibinfo  {journal} {Proc Natl Acad Sci U S A}\ }\textbf {\bibinfo {volume} {112}},\ \bibinfo {pages} {7183--8} (\bibinfo {year} {2015})}\BibitemShut {NoStop}%
\bibitem [{\citenamefont {Skinner}\ and\ \citenamefont {Dunkel}(2021)}]{SkinnerPRL}%
  \BibitemOpen
  \bibfield  {author} {\bibinfo {author} {\bibfnamefont {D.~J.}\ \bibnamefont {Skinner}}\ and\ \bibinfo {author} {\bibfnamefont {J.}~\bibnamefont {Dunkel}},\ }\bibfield  {title} {\enquote {\bibinfo {title} {Estimating entropy production from waiting time distributions},}\ }\href {\doibase 10.1103/PhysRevLett.127.198101} {\bibfield  {journal} {\bibinfo  {journal} {Phys. Rev. Lett.}\ }\textbf {\bibinfo {volume} {127}},\ \bibinfo {pages} {198101} (\bibinfo {year} {2021})}\BibitemShut {NoStop}%
\bibitem [{\citenamefont {Kampen}(2007)}]{van_Kampen_book}%
  \BibitemOpen
  \bibfield  {author} {\bibinfo {author} {\bibfnamefont {N.~G.~v.}\ \bibnamefont {Kampen}},\ }\href@noop {} {\emph {\bibinfo {title} {Stochastic processes in physics and chemistry}}},\ \bibinfo {edition} {3rd}\ ed.,\ North-Holland personal library\ (\bibinfo  {publisher} {Elsevier},\ \bibinfo {address} {Amsterdam ; Boston},\ \bibinfo {year} {2007})\ pp.\ \bibinfo {pages} {xvi, 463 p.}\BibitemShut {Stop}%
\end{thebibliography}%

\end{document}